\documentclass[runningheads]{llncs}

\usepackage{fontawesome}  
\usepackage{amsmath}
\usepackage{multirow}
\usepackage[table,xcdraw]{xcolor}
\usepackage{graphicx}
\usepackage{textcomp}
\usepackage{subfig}
\usepackage[utf8x]{inputenc}
\usepackage{csquotes}
\usepackage{flexisym}
\newcommand{\ignore}[1]{}
\usepackage[hidelinks]{hyperref}
\usepackage[font=scriptsize]{caption}

\usepackage[ruled,vlined,linesnumbered]{algorithm2e}

\usepackage{amssymb}
\usepackage[bottom]{footmisc}
\usepackage{wrapfig}
\usepackage{nicefrac}
\usepackage{bm}
\usepackage{mathrsfs}
\usepackage{xcolor}
\usepackage{hyperref}
\newcommand{\pactivities}{\mathcal{A}}

\newcommand{\presources}{\mathcal{R}}
\newcommand{\pcases}{\mathcal{C}}
\newcommand{\psensitive}{\mathcal{S}}
\newcommand{\processinstance}{\mathcal{P}}

\newcommand{\ptimes}{\mathcal{T}}
\newcommand{\pothers}{\mathcal{D}}

\newcommand{\RN}[1]{%
	\textup{\uppercase\expandafter{\romannumeral#1}}%
}
\newcommand{\spec}{\preceq_{n}}
\newcommand{\pinsprime}{(c',\sigma',s')}
\newcommand{\pins}{(c,\sigma,s)}
\newcommand{\pinsone}{(c_1,\sigma_1,s_1)}
\newcommand{\pinstwo}{(c_2,\sigma_2,s_2)}
\newcommand{\compare}{\overset{n}{\sim}}

\usepackage{mathtools}

\newcommand{\orcid}[1]{
	\href{https://orcid.org/#1}{\includegraphics[scale=0.4]{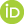}}
}

\newtheorem{exmp}{Example}

\begin{document}
\title{Privacy-Preserving Continuous Event Data Publishing}
\titlerunning{Privacy-Preserving Continuous Event Data Publishing}
%
%

\author{Majid Rafiei\orcid{0000-0001-7161-6927}\textsuperscript{\href{mailto:majid.rafiei@pads.rwth-aachen.de}{\faEnvelopeO}} \and
	Wil M.P. van der Aalst\orcid{0000-0002-0955-6940}}

\authorrunning{Majid Rafiei and Wil M.P. van der Aalst}
%
%
\institute{Chair of Process and Data Science, RWTH Aachen University, Aachen, Germany \\
 }
\maketitle              
\vspace{-0.3cm}
\begin{abstract}
Process mining enables organizations to discover and analyze their actual processes using event data.
Event data can be extracted from any information system supporting operational processes, e.g., SAP. Whereas the data inside such systems is protected using access control mechanisms, the extracted event data contain sensitive information that needs to be protected.
This creates a new risk and a possible inhibitor for applying process mining.
Therefore, privacy issues in process mining become increasingly important.
Several privacy preservation techniques have been introduced to mitigate possible attacks against static event data published only once. 
However, to keep the process mining results up-to-date, event data need to be published continuously. 
For example, a new log is created at the end of each week.
In this paper, we elaborate on the attacks which can be launched against continuously publishing anonymized event data by comparing different releases, so-called \textit{correspondence attacks}. Particularly, we focus on group-based privacy preservation techniques and show that provided privacy requirements can be degraded exploiting correspondence attacks. We apply the continuous event data publishing scenario to existing real-life event logs and report the anonymity indicators before and after launching the attacks.

\keywords{Process mining \and Privacy preservation \and Correspondence attacks \and Event data}

\end{abstract}
\section{Introduction}\label{sec:introduction}

Process mining bridges the gap between \textit{data science} and \textit{process science} using event logs. 
Event logs are widely available in different types of information systems \cite{van2016process}. Events are the smallest units of process execution which are characterized by their attributes. Process mining requires that each event contains at least the following main attributes to enable the application of analysis techniques: \textit{case id}, \textit{activity}, and \textit{timestamp}. 
The \textit{case id} refers to the entity that the event(s) belongs to, and it is considered as a process instance. The \textit{activity} refers to the activity associated with the event, and the \textit{timestamp} is the exact time when the activity was executed for the case. Moreover, depending on the context of a process, the corresponding events may contain more attributes. 
Table~\ref{tbl:sample_evenlog} shows a part of an event log recorded by an information system in a hospital.

In Table~\ref{tbl:sample_evenlog}, each row represents an event. A sequence of events, associated with a \textit{case id} and ordered using the timestamps, is called a \textit{trace}. 
Table~\ref{tbl:sample_evenlog_simple} shows a simple trace representation of Table~\ref{tbl:sample_evenlog} where the \textit{trace} attribute is a sequence of activities.  
Some of the event attributes may refer to individuals, e.g., the \textit{case id} refers to the patient whose data is recorded, and the \textit{resource} refers to the employees performing activities for the patients, e.g., surgeons. 
Also, some sensitive information may be included, e.g., the \textit{disease} attribute in Table~\ref{tbl:sample_evenlog}.
When individuals' data are included in an event log, privacy issues emerge, and organizations are obliged to consider such issues according to regulations, e.g., the European General Data Protection Regulation (GDPR)\footnote{http://data.europa.eu/eli/reg/2016/679/oj}.


\begin{table}[t]
	\parbox{.64\linewidth}{
		\centering
		\tiny
		\caption{Sample event log (each row represents an event).}\label{tbl:sample_evenlog}
		\begin{tabular}{ccccc}
			\hline
			Case Id & Activity          & Timestamp           & Resource  & Disease \\ \hline
			1       & Registration (RE) & 01.01.2019-08:30:00 & Employee1 & Flu     \\
			1       & Visit (VI)        & 01.01.2019-08:45:00 & Doctor1   & Flu     \\
			2       & Registration (RE) & 01.01.2019-08:46:00 & Employee1 & Corona  \\
			3       & Registration (RE) & 01.01.2019-08:50:00 & Employee1 & Cancer  \\
			...     & ...               & ...                 & ...       & ...     \\
			1       & Release (RL)      & 01.01.2019-08:58:00 & Employee2 & Flu     \\
			3       & Visit (VI)        & 01.02.2019-10:15:00 & Doctor3   & Cancer  \\
			2       & Release (RL)      & 01.02.2019-14:00:00 & Employee2 & Corona  \\
			3       & Blood Test (BT)   & 01.02.2019-14:15:00 & Employee5 & Cancer  \\
			...     & ...               & ...                 & ...       & ...     \\ \hline
		\end{tabular}
		
	}
	\hfill
	\parbox{.35\linewidth}{
		\centering
		\tiny
		\caption{A simple event log derived from Table~\ref{tbl:sample_evenlog} (each row represents a simple process instance).}\label{tbl:sample_evenlog_simple}
		\begin{tabular}{ccc}
			\hline
			Case Id & Trace                     & Disease \\ \hline
			1      & $\langle RE,VI,...,RL \rangle$ & Flu  \\
			2      & $\langle RE,...,RL \rangle$ & Corona     \\
			3      & $\langle RE,...,VI,BT,... \rangle$   & Cancer    \\
			...      & ...   & ...   \\ \hline
		\end{tabular}
	}
\end{table}

The privacy/confidentiality issues in process mining are recently receiving more attention. Various techniques have been proposed covering different aspects, e.g., confidentiality frameworks \cite{rafieiWA19_short}, privacy guarantees \cite{pretsaICPM2019_short,rafieitlkc_short,MannhardtKBWM19_short}, inter-organizational privacy issues \cite{smcProcessMining_short}, privacy quantification \cite{riskProcessMining_short,rafiei_quantification}, etc. 
Each of these approaches considers a single event log shared at some point in time.
This even log is published considering the privacy/confidentiality issues of a single log in isolation. 
However, event logs are recorded continuously and need to be published continuously to keep the results of process mining techniques updated.      

Continuous event data publishing lets an adversary launch new types of attacks that are impossible when event data are published only once. In this paper, we analyze the so-called \textit{correspondence attacks} \cite{fungCDP} that an adversary can launch by comparing different releases of anonymized event logs when they are continuously published. 
Particularly, we focus on group-based Privacy Preservation Techniques (PPTs) and describe three main types of correspondence attacks including \textit{forward attack}, \textit{cross attack}, and \textit{backward attack}. 
We analyze the privacy/anonymity losses imposed by these attacks and show how to detect such privacy losses efficiently. The explained anonymity analyses could be attached to different PPTs to empower them against the attacks or to change the data publishing approaches to bound such attacks. We applied different continuous event data publishing scenarios to several real-life event logs and report the anonymity indicators before and after launching the attacks for an example event log.        

The remainder of the paper is organized as follows. In Section~\ref{sec:motivation}, we present the problem statement. In Section~\ref{sec:prelimineries}, the preliminaries are explained. Different types of correspondence attacks are analyzed in Section~\ref{sec:attack_analysis}. In Section~\ref{sec:attack_detection}, we explain the attack detection techniques and privacy loss quantification. Section~\ref{sec:experiments} presents the experiments. Section~\ref{sec:extensions} discusses different aspects to extend the approach. Section~\ref{sec:related_work} discusses related work, and Section~\ref{sec:conclusion} concludes the paper.

%

\section{Problem Statement}\label{sec:motivation}
\begin{figure}[t]
	\centering
	\includegraphics[width=0.90\textwidth]{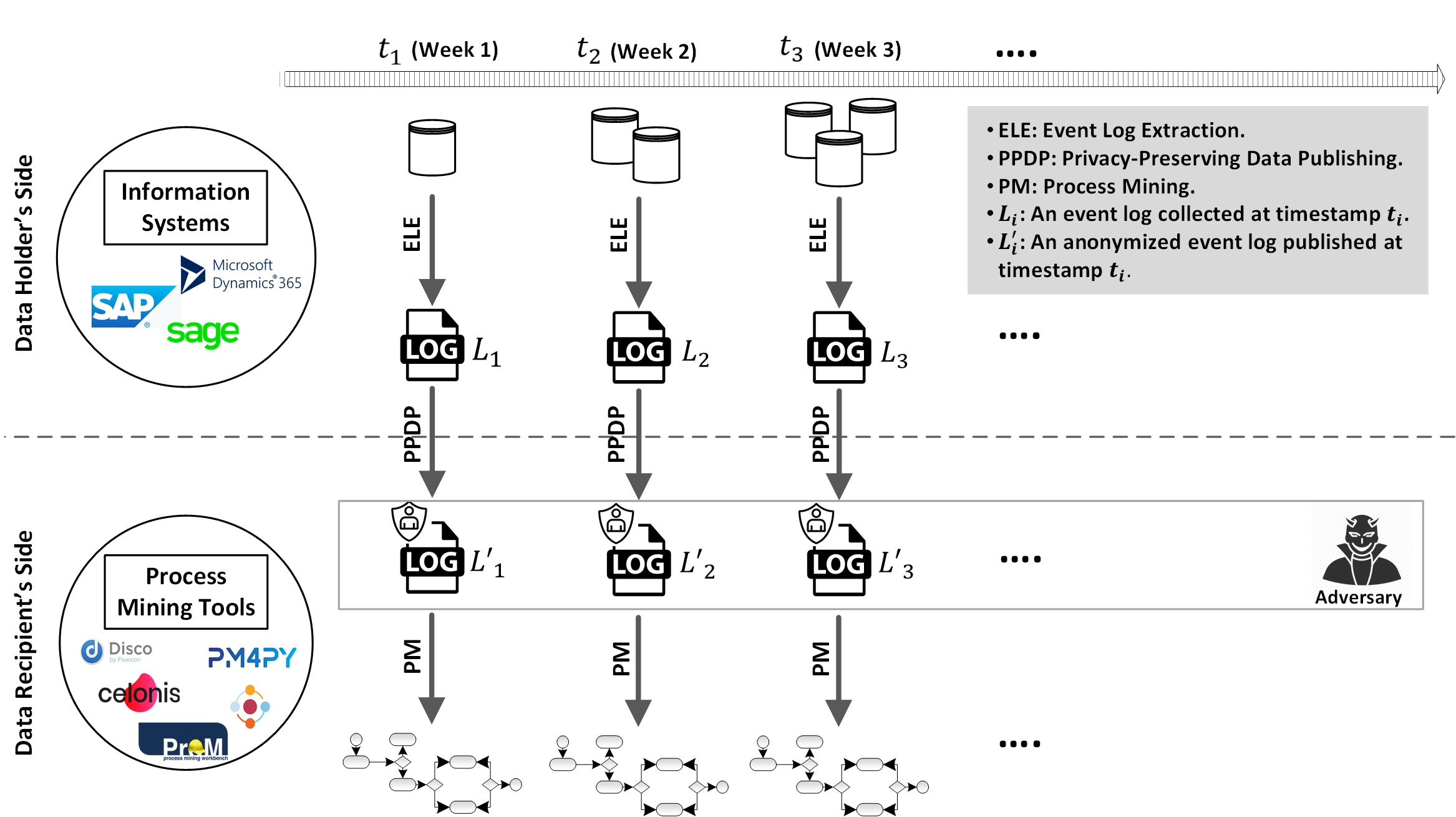}
	\caption{The general data collection and publishing scenario.}\label{fig:data_scenario}
\end{figure}

Figure~\ref{fig:data_scenario} shows our general data collection and publishing scenario. 
Information systems, e.g., SAP, provide operational support for organizations and continuously generate a lot of valuable event data. Such data are continuously collected and published, e.g., weekly, to be used by process mining tools, e.g., ProM, Disco, etc. On the analysis side, process mining techniques are applied to event logs to discover and analyze real processes supported by operational information systems. 
With respect to the types of data holder's models, introduced in \cite{Gehrke06_short}, we consider a \textit{trusted model} where the \textit{data holder}, i.e., the business owner, is trustworthy, but the \textit{data recipient}, i.e., a process miner, is not trustworthy. Therefore, PPTs are applied to event logs when they are published.

Continuous data publishing is generally classified into three main categories: \textit{incremental}, \textit{decremental}, and \textit{dynamic} \cite{fung2010introduction}.
Continuous event data publishing is considered as \textit{incremental}, i.e., the events generated by an information system are cumulatively collected, and they are not updated or deleted after the collection.
Thus, the so-called \textit{correspondence knowledge} is gained.
If we assume that in a continuous event data publishing scenario, the event logs are collected and published weekly, the correspondence knowledge is as follows:
(1) Every case started in the $i$-th week is in the $i$-th event log $L_i$, and must be in $L_j$, $i{<}j$, and (2) Every case started in the $j$-th week is in the $j$-th event log $L_j$, and cannot be in $L_i$, $i{<}j$.
Although each single anonymized event log $L'$ meets the privacy guarantees specified in the corresponding PPT, the adversary, who has access to the different releases of anonymized event logs, can exploit the \textit{correspondence knowledge} to degrade the provided privacy guarantees.  

Consider Table~\ref{tbl:r1_motivation} and Table~\ref{tbl:r2_motivation} as two anonymized event logs, $L_1'$ and $L_2'$, published at timestamps $t_1$ (week 1) and $t_2$ (week 2), respectively.
Note that the case identifiers are dummy identifiers independently assigned to the cases of each release. 
If we assume that an adversary's Background Knowledge (BK) is a sequence of activities with maximum length 3, both published event logs have 2-anonymity and 2-diversity. 
Assume the situation where the adversary knows that $\langle a,b,c \rangle$ is a subsequence of activities performed for a victim case, and that the process of the case has been started in the second week, i.e., it should be included in Table~\ref{tbl:r2_motivation}. Based on the correspondence knowledge, the only matching case is 30. 
Note that by a simple comparison of $L_1'$ and $L_2'$ based on the \textit{disease} attribute, it is obvious that cases 10 and 20 have to be started in the first week and cannot match the adversary's BK. This is called \textit{backward attack} ($B$-attack) which is a specific type of the correspondence attacks.

\begin{table}[t]
	\parbox{.49\linewidth}{
		\centering
		\tiny
		\caption{An anonymized event log published at timestamp $t_1$ (e.g., week 1), meeting 2-anonymity and 2-diversity when the assumed BK is a sequence of activities with the maximum length 3.}\label{tbl:r1_motivation}
		\begin{tabular}{ccc}
			\hline
			Case Id & Trace                     & Disease \\ \hline
			1       & $\langle a,b,c,d \rangle$ & Corona  \\
			2       & $\langle a,b,c,d \rangle$ & Flu     \\
			3       & $\langle a,e,d \rangle$   & Fever   \\
			4       & $\langle a,e,d \rangle$   & Corona  \\ \hline
		\end{tabular}
		
	}
	\hfill
	\parbox{.49\linewidth}{
		\centering
		\tiny
		\caption{An anonymized event log published at timestamp $t_2$ (e.g., week 2), meeting 2-anonymity and 2-diversity when the assumed BK is a sequence of activities with the maximum length 3.}\label{tbl:r2_motivation}
		\begin{tabular}{ccc}
			\hline
			Case Id & Trace                     & Disease \\ \hline
			10      & $\langle a,b,c,d \rangle$ & Corona  \\
			20      & $\langle a,b,c,d \rangle$ & Flu     \\
			30      & $\langle a,b,c,d \rangle$   & HIV     \\
			40      & $\langle a,e,d \rangle$   & Fever   \\
			50      & $\langle a,e,d \rangle$   & Corona  \\ \hline
		\end{tabular}
	}
\end{table}

The provided attack scenario shows that when event logs are collected and published continuously, the corresponding PPDP approaches need to be equipped with some techniques to detect the potential attacks that can be launched by an adversary who receives various anonymized event logs.
In this paper, we focus on simple event logs and group-based PPTs, i.e., $k$-anonymity, $l$-diversity, $t$-closeness, etc. 
We first describe the approach based on two releases of event logs, then we explain the possible extensions for any number of releases. 


\section{Preliminaries}\label{sec:prelimineries}
We first introduce some basic notations. For a given set $A$, $A^*$ is the set of all finite sequences over $A$. A finite sequence over $A$ of length $n$ is a mapping $\sigma {\in} \{1,...,n\} \rightarrow{A}$, represented as $\sigma {=} \langle a_1,a_2,...,a_n \rangle$ where $a_i {=} \sigma(i)$ for any $1{\leq} i {\leq} n$. $|\sigma|$ denotes the length of the sequence. 
For $\sigma_1, \sigma_2 {\in} A^*$, $\sigma_1 {\sqsubseteq} \sigma_2$ if $\sigma_1$ is a subsequence of $\sigma_2$, e.g., $\langle z,b,c,x \rangle {\sqsubseteq} \langle z,x,a,b,b,c,a,b,c,x \rangle$.
For $\sigma=\langle a_1,a_2,...,a_n \rangle$, $pref(\sigma){=}\{\langle a_1,...,a_k \rangle \mid 1 {\le} k {\le} n \}$, e.g., $\langle a,b,c,d \rangle \in pref(\langle a,b,c,d,e,f \rangle)$. 

\begin{definition}[LCS and SCS]
	\label{def:lcs_SCS}
	Let $\sigma_1 \in A^*$ and $\sigma_2 \in A^*$ be two sequences. $CSB(\sigma_1,\sigma_2){=}\{ \sigma {\in} A^* \mid \sigma {\sqsubseteq} \sigma_1 \wedge \sigma {\sqsubseteq} \sigma_2\}$ is the set of common subsequences, and $LCS(\sigma_1,\sigma_2) {=} \{ \sigma {\in} CSB \mid  \forall_{\sigma' {\in} CSB(\sigma_1,\sigma_2)}|\sigma'| {\le} |\sigma| \}$ is the set of longest common subsequences.
	$LCS^{\sigma_1}_{\sigma_2}$ denotes the length of a longest common subsequence for $\sigma_1$ and $\sigma_2$.
	Also, $CSP(\sigma_1,\sigma_2){=}\{ \sigma {\in} A^* \mid \sigma_1 {\sqsubseteq} \sigma \wedge \sigma_2 {\sqsubseteq} \sigma\}$ is the set of common super-sequences, and $SCS(\sigma_1,\sigma_2) {=}\{ \sigma {\in} CSB \mid  \forall_{\sigma' {\in} CSP(\sigma_1,\sigma_2)}|\sigma'| {\ge} |\sigma| \}$ is the set of shortest common super-sequences.
	$SCS^{\sigma_1}_{\sigma_2}$ denotes the length of a shortest common super-sequence for $\sigma_1$ and $\sigma_2$.
\end{definition}

\begin{definition}[Event, Event Log]
	\label{def:event}
	An event is a tuple $e = (c,a,t,r,d_1,...,d_m)$, where $c {\in} \pcases$ is the \textit{case id}, $a {\in} \pactivities$ is the activity associated with the event, $t {\in} \ptimes$ is the event timestamp, $r {\in} \presources$ is the \textit{resource}, who is performing the activity, and $d_1$,...,$d_m$ is a list of additional attributes values, where for any $1\leq i \leq m, d_i {\in} \pothers_i$. We call $\xi = \pcases {\times} \pactivities {\times} \ptimes {\times} \presources {\times} \pothers_1 {\times} ... {\times} \pothers_m$ the event universe.
	For $e = (c,a,t,r,d_1,...,d_m)$, $\pi_c(e){=}c$, $\pi_a(e){=}a$, $\pi_t(e){=}t$, $\pi_r(e){=}r$, and $\pi_{d_i}(e){=}d_i$, $1{\leq} i {\leq} m$, are its projections.
	An \textbf{event log} is $L {\subseteq} \xi$ where events are unique.
\end{definition}

In continuous event data publishing, event logs are collected and published continuously at each timestamp $t_i$, $i {\in} \mathbb{N}_{\geq 1}$. $L_i$ is the event log collected at the timestamp $t_i$, i.e., $L_i=\{ e {\in} \xi \mid \pi_{t}(e) {\leq} t_i \}$.
For $L_i$ and $L_j$, s.t., $i{<}j$, $L_j$ could contain new events for the cases already observed in $L_i$ and new cases not observed in $L_i$.
In the following, we define a simple version of event logs which will later be used for demonstrating the attacks and corresponding anonymity measures. 

\begin{definition}[Trace, Simple Trace]
	\label{def:trace}
	A trace $\sigma{=}\langle e_1,e_2,...,e_n \rangle {\in} \xi^*$ is a sequence of events, s.t., for each $e_i,e_j {\in} \sigma$: $\pi_{c}(e_i){=}\pi_{c}(e_j)$, and $\pi_{t}(e_i) {\le} \pi_{t}(e_j)$ if $i {<} j$.
	A \textit{simple trace} is a trace where all the events are projected on the activity attribute, i.e., $\sigma \in \pactivities^*$.
\end{definition}

\begin{definition}[Simple Process Instance]
		\label{def:simplePI}
		We define $\processinstance {=} \pcases {\times} \pactivities^* {\times} \psensitive$ as the universe of simple process instances, where $\psensitive {\subseteq} \pothers_1 {\cup} ... {\cup} \pothers_m$ is the domain of the sensitive attribute. Each simple process instance $\pins {\in} \processinstance$ represents a \textbf{simple trace} $\sigma {=} \langle a_1,a_2,...,a_n \rangle$, belonging to the case $c$ with $s$ as the sensitive attribute value.
		For $p{=}\pins {\in} \processinstance$, $\pi_c(p){=}c$, $\pi_\sigma(p){=}\sigma$, and $\pi_s(p){=}s$ are its projections.
		
\end{definition}


\begin{definition}[Simple Event Log]
		\label{def:simpleEL}
		 Let $\processinstance {=} \pcases {\times} \pactivities^* {\times} \psensitive$ be the universe of simple process instances. A simple event log is $L {\subseteq} \processinstance$, s.t., if ${\pinsone \in L}$, $\pinstwo \in L$, and $c_1{=}c_2$, then $\sigma_1 {=} \sigma_2$ and $s_1{=}s_2$. 
\end{definition}


\section{Attack Analysis}\label{sec:attack_analysis}
We analyze the correspondence attacks by focusing on two anonymized releases obtained by applying group-based PPTs to simple event logs. 
In general, group-based PPTs provide desired privacy requirements utilizing \textit{suppression} and/or \textit{generalization} operations. Particularly, the group-based PPTs introduced for the event data protection are mainly based on the \textit{suppression} operation \cite{pretsaICPM2019_short,rafieitlkc_short}, where some events are removed to provide the desired privacy requirements. Hence, apart from any specific privacy preservation algorithm, we define a general anonymization function that converts an event log to another one meeting desired privacy requirements assuming a bound for the maximum number of events that can be removed from each trace, so-called the \textit{anonymization parameter}. Note that this assumption is based on the \textit{minimality principle} in PPDP \cite{minimality}. Similar attack analysis can be done for the generalization operation as well. 

\begin{definition}[Anonymization]
	\label{def:anon}
	Let $\processinstance$ be the universe of simple process instances and $n {\in} \mathbb{N}_{\ge 1}$ be the anonymization parameter. We define $anon^n \in 2^\processinstance \rightarrow{2^\processinstance}$ as a function for anonymizing event logs.
	For all $L,L'{ \subseteq} \processinstance$, $anon^n(L){=}L'$ if there exists a bijective function $f {\in} L \rightarrow{L'}$, s.t., for any $p{=}\pins {\in} L$ and $p'{=}\pinsprime {\in} L'$ with $f(p){=}p'{:}$ $\sigma'{\sqsubseteq}\sigma$, $|\sigma|-n {\leq} |\sigma'|$, and $s'{=}s$.  
\end{definition}

\begin{figure}[t]
	\centering
	\includegraphics[width=0.98\textwidth]{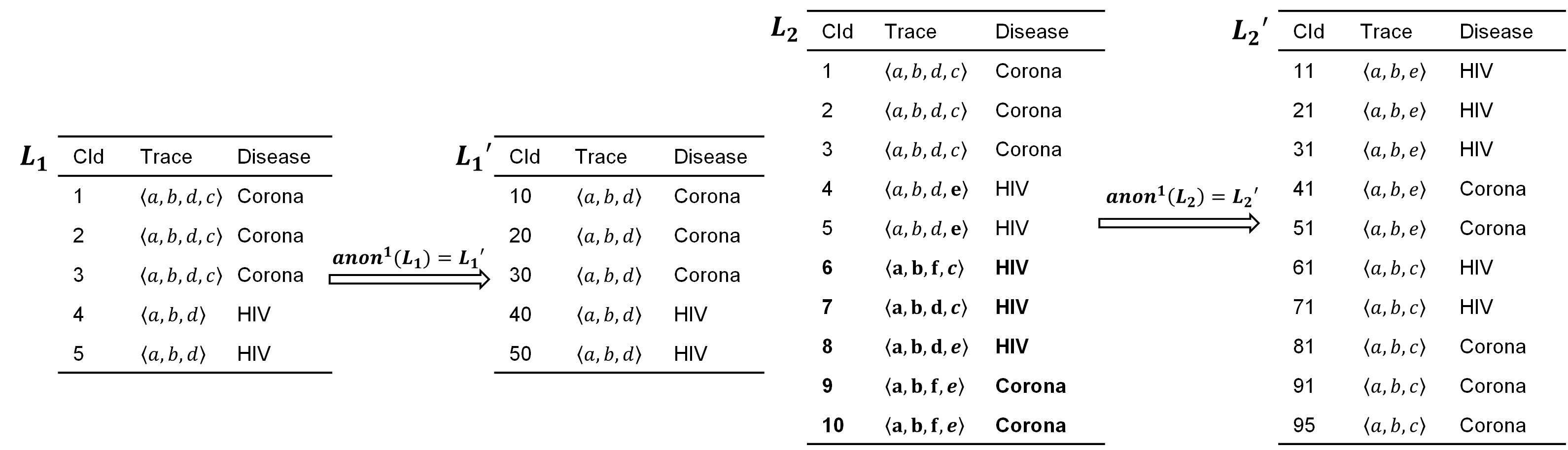}
	\caption{$L_1$ and $L_2$ are two simple event logs collected at timestamps $t_1$ and $t_2$. $L_1'$ and $L_2'$ are the corresponding anonymized releases of event logs given $n{=}1$ as the anonymization parameter. Both $L_1'$ and $L_2'$ have 5-anonymity and 2-diversity assuming a sequence of activities as the BK.}\label{fig:anon}
\end{figure}

Note that we assume the anonymization function promises to preserve all the cases and not to produce new (fake) cases. Figure~\ref{fig:anon} shows two simple event logs that were published using the anonymization function given $n=1$. \textit{Specialization} is the reverse operation of the anonymization defined as follows.

\begin{definition}[Specialization]
	\label{def:sp}
	Let $\processinstance$ be the universe of simple process instances and $n {\in} \mathbb{N}_{\ge 1}$ be the anonymization parameter. 
	For $p{=}\pins {\in} \processinstance$ and $p'{=}\pinsprime {\in} \processinstance$, we say $p$ is a specialization for $p'$ w.r.t. $n$, denoted by $p' {\spec} p$ iff $\sigma'{\sqsubseteq}\sigma$, $|\sigma| {\leq} |\sigma'| {+}n$, and $s=s'$.
\end{definition}

Consider $p'{=}(81,\langle a,b,c \rangle, Corona)$ as a process instance from the anonymized event log $L_2'$ in Figure~\ref{fig:anon}. Given $n{=}1$, the cases 1, 2, and 3 from $L_2$ could be a specialization for $p'$ which are possible original process instances. 
We assume that the adversary's BK is a subsequence of activities performed for a victim case which can be considered as the strongest assumable knowledge w.r.t. the available information in simple event logs.
Given an anonymized event log and the anonymization parameter, the adversary can distinguish a \textit{matching set} in the anonymized release containing all the process instances having at least one specialization matching the adversary's knowledge. One of the process instances included in such a matching set belongs to the victim case.

\begin{definition}[Matching Set, Group]
	\label{def:match}
	Let $n {\in} \mathbb{N}_{\ge 1}$ be the anonymization parameter and $L'$ be an anonymized event log. $ms^{L',n} {\in} \pactivities^* \rightarrow{2^{L'}}$ retrieves a set of matching process instances from $L'$.
	For $bk {\in} \pactivities^*$, $ms^{L',n}(bk){=}\{ p' {\in} L' \mid \exists_{p {\in} \processinstance} p' {\spec} p \wedge bk {\sqsubseteq} \pi_\sigma(p) \}$. A \textbf{group} $g$ in a matching set is a set of process instances having the same value on the sensitive attribute.
\end{definition}

Consider $bk {=} \langle d,e \rangle$ as the adversary's knowledge and $n{=}1$. For the anonymized event logs in Figure~\ref{fig:anon}, $ms^{L_1',n}(bk){=} L_1'$, and $ms^{L_2',n}(bk) {=} \{ (c',\sigma',s') {\in} L_2' \mid c' {\in} \{ 11,21\\,31,41,51 \}\}$.
The elements of matching sets can be identified using the following theorem without searching the space of specializations.

\begin{theorem}[Elements of matching sets]
	\label{theory:match}
	Let $n {\in} \mathbb{N}_{\ge 1}$ be the anonymization parameter and $L'$ be an anonymized event log. For $bk {\in} \pactivities^*$ and $p'{=}(c',\sigma',s') {\in} L'$, $p' {\in} ms^{L',n}(bk)$ iff $n \ge |bk| - LCS^{bk}_{\sigma'}$.
	
\end{theorem}
\vspace{-0.6 cm}
\begin{proof}
	Theorem~\ref{theory:match} follows because one needs to add at least $|bk| - LCS^{bk}_{\sigma'}$ activities to generate a super-sequence $\sigma$ of $\sigma'$, s.t., ${bk \sqsubseteq \sigma}$. $\sigma$ can be considered as the trace of a process instance $p$ which is a specialization for $p'$. Note that one can always assign a value for the sensitive attribute of $p'$, s.t., $\pi_s(p){=}\pi_s(p')$. 
\end{proof}

Consider a scenario where the data holder publishes $L_1'$ and $L_2'$ as two anonymized event logs at timestamps $t_1$ and $t_2$, respectively. An adversary, who is one of the data recipients, attempts to identify a victim case $vc$ from $L_1'$ or $L_2'$. We assume that the adversary's knowledge is a subsequence of activities performed for the $vc$, i.e., $bk \in \pactivities^*$, and the approximate time at which the process of the $vc$ has been started, which is enough to know the release(s) where the $vc$ should appear. For example, if event logs are published weekly, then the adversary knows that the process of the $vc$ has been started in the second week. Thus, its data should appear in all the event logs published after the first week. The adversary has also the \textit{correspondence knowledge} derived from the concept of continuous event data publishing, as described in Section~\ref{sec:motivation}.
The following correspondence attacks can be launched by the adversary.       

\textbf{Forward Attack ($F$-attack)}
The adversary knows that the process of the $vc$ has been started at the approximate time $t$, s.t., $t{\le}t_1$, and tries to identify the $vc$ in $L_1'$ exploiting $L_2'$ and $bk {\in} \pactivities^*$ as the BK. 
The $vc$ due to its timestamp must have a process instance in $L_1'$ and $L_2'$. 
If there exists a $p_1' {\in} L_1'$, s.t., $p_1' {\in} ms^{L_1',n}(bk)$ for an anonymization parameter $n$, there must be a $p_2' {\in} L_2'$ corresponding to $p_1'$. Otherwise, $p_1'$ does not match the BK and can be excluded from $ms^{L_1',n}(bk)$. 

\begin{exmp}
	Consider $L_1'$ and $L_2'$ in Figure~$\ref{fig:anon}$. 
	Assume that the adversary's knowledge is $bk{=}\langle d,e \rangle$, and the anonymization parameter is $n{=}1$.
	$ms^{L_1',n}(bk){=} L_1'$ and $ms^{L_2',n}(bk) {=} \{ \pinsprime {\in} L_2' \mid c' {\in} \{ 11,21,31,41,51 \}\}$. Both matching sets meet $5$-anonymity. However, by comparing $L_1'$ and $L_2'$, the adversary learns that one of the cases $10,20,30$ cannot have $e$ after $d$. Otherwise, there must have been three cases with Corona in $ms^{L_2',n}(bk)$. Therefore, the adversary can exclude one of $10,20,30$. Note that the choice among $10,20,30$ does not matter as they are equal. Consequently, $k$ is degraded from $5$ to $4$.    
\end{exmp}

\textbf{Cross Attack ($C$-attack)}
The adversary knows that the process of the $vc$ has been started at the approximate time $t$, s.t., $t{\le}t_1$, and attempts to identify the $vc$ in $L_2'$ exploiting $L_1'$ and $bk {\in} \pactivities^*$ as the BK. 
The $vc$ because of its timestamp must have a process instance in $L_1'$ and $L_2'$.
If there exists a $p_2' {\in} L_2'$, s.t., $p_2' {\in} ms^{L_2',n}(bk)$ for an anonymization parameter $n$, there must be a $p_1' {\in} L_1'$ corresponding to $p_2'$.
Otherwise, $p_2'$ either is started at timestamp $t$, $t_1{<}t{\leq}t_2$, or it does not match the BK and can be excluded from $ms^{L_2',n}(bk)$. 

\begin{exmp}
	Consider $L_1'$ and $L_2'$ in Figure~$\ref{fig:anon}$. 
	Assume that the adversary's knowledge is $bk{=}\langle d,e \rangle$, and the anonymization parameter is $n{=}1$.
	$ms^{L_1',n}(bk)= L_1'$ and $ms^{L_2',n}(bk) {=} \{ \pinsprime {\in} L_2' \mid c' {\in} \{ 11,21,31,41,51 \}\}$. Both matching sets meet $5$-anonymity. However, by comparing $L_1'$ and $L_2'$, the adversary learns that one of the cases $11,21,31$ must be started at timestamp $t$, s.t., $t_1{<}t{\le}t_2$. Otherwise, there must have been three cases with HIV in $ms^{L_1',n}(bk)$. Therefore, the adversary can exclude one of $11,21,31$. Again, the choice among $11,21,31$ does not matter as they are equal. Consequently, $k$ is degraded from $5$ to $4$.   
\end{exmp}     

\textbf{Backward Attack ($B$-attack)}
The adversary knows that the process of the $vc$ has been started at the approximate time $t$, s.t., $t_1{<}t{\le}t_2$, and tries to identify the $vc$ in $L_2'$ exploiting $L_1'$ and $bk {\in} \pactivities^*$ as the BK. 
The $vc$ has a process instance in $L_2'$, but not in $L_1'$. Hence, if there exists $p_2' \in L_2'$, s.t., $p_2' \in ms^{L_2',n}(bk)$ for an anonymization parameter $n$, and $p_2'$ has to be a corresponding process instance for some process instances in $L_1'$, then $p_2'$ must be started at timestamp $t$, s.t., $t{\le}t_1$ and can be excluded from the matching set $ms^{L_2',n}(bk)$.
      
\begin{exmp}
	Consider $L_1'$ and $L_2'$ in Figure~$\ref{fig:anon}$.
	Assume that the adversary's knowledge is $bk{=}\langle d,c \rangle$, and the anonymization parameter is $n{=}1$.
	$ms^{L_1',n}(bk){=} L_1'$ and $ms^{L_2',n}(bk) {=} \{ \pinsprime {\in} L_2' \mid c' {\in} \{ 61,71,81,91,95 \}\}$. 
	Both matching sets meet $5$-anonymity. However, by comparing $L_1'$ and $L_2'$, the adversary learns that at least one of the cases $81,91,95$ must be started at timestamp $t$, $t{\le}t_1$. Otherwise, one of the cases $10,20,30$ cannot have a corresponding process instance in $L_2'$. Thus, $k$ is degraded from $5$ to $4$. Note that there are only two cases with Corona which are not in $ms^{L_2',n}(bk)$ and could be corresponding for cases $10,20,30$. Hence, at least one of $81,91,95$ must be started at timestamp $t$, $t{\le}t_1$. 
\end{exmp}
     
\section{Attack Detection}\label{sec:attack_detection}
The correspondence attacks mentioned in Section~\ref{sec:attack_analysis} are based on making some inferences about corresponding cases (process instances). However, there are many possible assignments of corresponding cases and each of those implies possibly different event logs, which are not necessarily the actual event logs collected by the data holder. In this section, we demonstrate the \textit{attack detection} regardless of any particular choices. To this end, we first need to define a \textit{linker} to specify all the valid assignments.
Then, we provide formal definitions for different types of correspondence attacks and corresponding anonymity indicators.

\begin{definition}[Linker, Buddy]
	\label{def:linker}
	Let $L_1'$ and $L_2'$ be the anonymized event logs at timestamps $t_1$ and $t_2$, respectively, and $n {\in} \mathbb{N}_{\ge 1}$ be the anonymization parameter. 
	$linker^n {\in} L_1' {\rightarrow}{L_2'}$ is a total injective function that retrieves the corresponding process instances.
	For $p_1' {\in} L_1'$ and $p_2' {\in} L_2'$, $linker^n(p_1'){=}p_2'$ iff there exist $p_1,p_2 {\in} \processinstance$, s.t., $p_1' {\spec} p_1 \wedge p_2' {\spec} p_2 \wedge \pi_s(p_1){=}\pi_s(p_2) \wedge  \pi_\sigma(p_1) \in pref(\pi_\sigma(p_2))$. $(p_1',p_2')$ is called a pair of \textbf{buddies} if there exists a linker, s.t., $linker^n(p_1'){=}p_2'$. 
\end{definition}

\begin{definition}[$F$-attack]
	\label{def:f-attack}
	Let $L_1'$ and $L_2'$ be two anonymized event logs released at timestamps $t_1$ and $t_2$, $n {\in} \mathbb{N}_{\ge 1}$ be the anonymization parameter, $t {\le} t_1$ be the approximate time at which the process of the victim case has been started, and $bk {\in} \pactivities^*$ be the BK. The $F$-attack attempts to identify $x$ as the maximal excludable cases from $ms^{L_1',n}(bk)$, s.t., for any linker, at least $x$ cases from the matching set cannot match the BK. $x$ is considered as \textit{crack size} based on $F$-attack. 
\end{definition}

\begin{definition}[$C$-attack]
	\label{def:c-attack}
	Let $L_1'$ and $L_2'$ be two anonymized event logs released at timestamps $t_1$ and $t_2$, $n {\in} \mathbb{N}_{\ge 1}$ be the anonymization parameter, $t {\le} t_1$ be the approximate time at which the process of the victim case has been started, and $bk {\in} \pactivities^*$ be the BK. The $C$-attack tries to identify $x$ (crack size) as the maximal excludable cases from $ms^{L_2',n}(bk)$, s.t., for any linker, at least $x$ cases from the matching set cannot match the BK or the timestamp of the victim case.
\end{definition}

\begin{definition}[$B$-attack]
	\label{def:b-attack}
	Let $L_1'$ and $L_2'$ be two anonymized event logs released at timestamps $t_1$ and $t_2$, $n {\in} \mathbb{N}_{\ge 1}$ be the anonymization parameter, $t_1{<} t{\le} t_2$ be the approximate time at which the process of the victim case has been started, and $bk {\in} \pactivities^*$ be the BK. The $B$-attack tries to identify $x$ (crack size) as the maximal excludable cases from $ms^{L_2',n}(bk)$, s.t., for any linker, at least $x$ cases from the matching set cannot match the timestamp of the victim case.
\end{definition}

Based on the definitions for the correspondence attacks, the key for attack detection is the crack size. For calculating the crack sizes, we follow the similar approach introduced in \cite{fungCDP} which is based on the concept of \textit{comparability}. We define the comparability at the level of \textit{sequences}, \textit{process instances}, and \textit{groups}. These definitions are later used to compute the crack sizes of attacks.

\begin{definition}[Comparable Sequences]
	\label{def:compare_seq}
	Let $\sigma_1, \sigma_2 {\in} \pactivities^*$ be two sequences of activities. We say $\sigma_1$ and $\sigma_2$ are comparable w.r.t. $n {\in} \mathbb{N}_{\ge 1}$, denoted by $\sigma_1 {\compare} \sigma_2$, if $n$ is the minimum number of activities that needs to be added to $\sigma_1$ and/or $\sigma_2$ to generate a joint super-sequence, or if $\sigma_1$ can be a prefix of $\sigma_2$ by adding at least $n$ activities to $\sigma_2$. 
\end{definition}

\begin{theorem}[Detecting comparable sequence]
	\label{theory:compare_seq}
	Given $\sigma_1, \sigma_2 {\in} \pactivities^*$ and $n {\in} \mathbb{N}_{\ge 1}{:}$
	\vspace{-0.18 cm}
	$
	\footnotesize
	\sigma_1 {\compare} \sigma_2  {\iff }
	\begin{cases}
	n \ge |\sigma_1| - LCS^{\sigma_1}_{\sigma_2}  & \text{if } \exists_{\sigma \in LCS(\sigma_1,\sigma_2)} \sigma {\in} pref(\sigma_2)\\
	n \ge SCS^{\sigma_1}_{\sigma_2} - min(|\sigma_1| ,|\sigma_2|) & \text{otherwise}
	\end{cases}       
	$
\end{theorem}
\vspace{-0.15 cm}
\begin{proof}
	If there exists a $\sigma {\in} LCS(\sigma_1,\sigma_2)$, s.t., $\sigma {\in} pref(\sigma_2)$, then $|\sigma_1| -  LCS^{\sigma_1}_{\sigma_2}$ is the minimum number of activities that needs to be added to $\sigma_2$, s.t., $\sigma_1 \in pref(\sigma_2)$. 
	Otherwise, since $SCS^{\sigma_1}_{\sigma_2}$ is the length of a shortest common super-sequence, one needs to add at least $SCS^{\sigma_1}_{\sigma_2} - min(|\sigma_1| ,|\sigma_2|)$ activities to the shorter sequence to generate a joint super-sequence.
\end{proof}


\begin{definition}[Comparable Process Instances]
	\label{def:compare_process}
	Let $p_1,p_2 {\in} \processinstance$ be two process instances. We say $p_1$ and $p_2$ are comparable w.r.t. $n$, denoted by $p_1 {\compare} p_2$, iff $\pi_s(p_1){=}\pi_s(p_2) \wedge \pi_\sigma(p_1) {\compare} \pi_\sigma(p_2)$.
\end{definition}


\begin{definition}[Comparable Groups]
	\label{def:compare_group}
	Let $L_1'$ and $L_2'$ be two anonymized event logs released at timestamps $t_1$ and $t_2$, $bk \in \pactivities^*$ be the BK, and $n {\in} \mathbb{N}_{\ge 1}$ be the anonymization parameter. 
	We say two groups $g_1' {\subseteq} ms^{L_1',n}(bk)$ and $g_2' {\subseteq} ms^{L_2',n}(bk)$ are comparable w.r.t. $n$, denoted by $g_1' {\compare} g_2'$, iff ${\forall_{p_1' \in g_1'}\forall_{p_2' \in g_2'} p_1' {\compare} p_2'}$. 
\end{definition}


\begin{lemma}
	\label{lemma:compare_2}
	Let $L_1'$ and $L_2'$ be two anonymized event logs at timestamps $t_1$ and $t_2$, $bk \in \pactivities^*$ be the BK, and $n {\in} \mathbb{N}_{\ge 1}$ be the anonymization parameter. 
	Consider $g_1' {\subseteq} ms^{L_1',n}(bk)$ and $g_2' {\subseteq} ms^{L_2',n}(bk)$ as two groups, s.t., $g_1' {\compare} g_2'$. If $p_1' {\in} ms^{L_1',n}(bk)$ and $p_2' {\in} ms^{L_2',n}(bk)$ are buddies for a linker, then $p_1' {\in} g_1'$ iff $p_2' {\in} g_2'$.  
\end{lemma}



\begin{lemma}
	\label{lemma:buddy_max}
	Let $L_1'$ and $L_2'$ be two anonymized event logs released at timestamps $t_1$ and $t_2$, $bk \in \pactivities^*$ be the BK, and $n {\in} \mathbb{N}_{\ge 1}$ be the anonymization parameter. 
	Consider $g_1' {\subseteq} ms^{L_1',n}(bk)$ and $g_2' {\subseteq} ms^{L_2',n}(bk)$ as two groups, s.t., $g_1' {\compare} g_2'$. Since the buddy relationship is injective, at most $min(|g_1'|,|g_2'|)$ process instances in $g_1'$ have a buddy in $g_2'$, and there are some linkers where exactly $min(|g_1'|,|g_2'|)$ process instances in $g_1'$ have a buddy in $g_2'$.
\end{lemma}

\begin{theorem}[Crack size based on $F$-attack]
	\label{theory:F_anonymity}
	Let $bk {\in} \pactivities^*$ be the BK, $n {\in} \mathbb{N}_{\ge 1}$ be the anonymization parameter, and $L_1'$ and $L_2'$ be two anonymized event logs released at timestamps $t_1$ and $t_2$. 
	Let $CG(ms^{L_1',n}(bk),ms^{L_2',n}(bk))=\{ (g_1',g_2') \mid g_1' {\subseteq} ms^{L_1',n}(bk) \wedge g_2' {\subseteq} ms^{L_2',n}(bk) \wedge g_1' {\compare} g_2' \}$ be the set of pair of comparable groups in the matching sets.
	For $(g_1',g_2') \in CG(ms^{L_1',n}(bk),ms^{L_2',n}(bk))$, $g_1'$ has crack size $cs=|g_1'|-min(|g_1'|,|g_2'|)$. 
	$F(ms^{L_1',n}(bk),ms^{L_2',n}(bk))=\sum cs$ is the number of excludable cases from $ms^{L_1',n}(bk)$ exploiting the $F$-attack, where $\sum$ is over $(g_1',g_2') {\in} CG(ms^{L_1',n}(bk),ms^{L_2',n}(bk))$.
\end{theorem}
\vspace{-0.5 cm}
\begin{proof}
	Consider $(g_1',g_2') {\in} CG(ms^{L_1',n}(bk),ms^{L_2',n}(bk))$. Based on \autoref{lemma:buddy_max}, if ${|g_1'| > |g_2'|}$, at least $|g_1'|-min(|g_1'|,|g_2'|)$ process instances in $g_1'$ do not have a buddy in $g_2'$ for any linker. Also, according to \autoref{lemma:compare_2}, these process instances cannot match the given BK. Otherwise, they must have had buddies in $g_2'$.  
\end{proof}

\begin{exmp}
	Consider $L_1'$ and $L_2'$ in Figure~$\ref{fig:anon}$, $n{=}1$, and $bk{=}\langle d,e \rangle$. $|g_1'|{=}3$ and $|g_2'|{=}2$ for the Corona groups in $ms^{L_1',n}(bk)$ and $ms^{L_2',n}(bk)$, respectively. $cs{=}3-min(3,2)$ is the crack size of $ms^{L_1',n}(bk)$ based on $F$-attack.
\end{exmp}

\begin{definition}[$F$-Anonymity]
	\label{def:f-anonymity}
	Let $L_1'$ and $L_2'$ be two anonymized event logs at $t_1$ and $t_2$, and $n {\in} \mathbb{N}_{\ge 1}$ be the anonymization parameter. The $F$-anonymity of $L_1'$ and $L_2'$ is ${FA}^{n}(L_1'{,}L_2')=\min\limits_{bk \in \pactivities^*}|ms^{L_1',n}(bk)|-F(ms^{L_1',n}(bk),ms^{L_2',n}(bk))$.
\end{definition}

\begin{theorem}[Crack size based on $C$-attack]
	\label{theory:C_anonymity}
	Let $bk {\in} \pactivities^*$ be the BK, $n {\in} \mathbb{N}_{\ge 1}$ be the anonymization parameter, and $L_1'$ and $L_2'$ be two anonymized event logs released at timestamps $t_1$ and $t_2$. 
	Let $CG(ms^{L_1',n}(bk),ms^{L_2',n}(bk))=\{ (g_1',g_2') \mid g_1' {\subseteq} ms^{L_1',n}(bk) \wedge g_2' {\subseteq} ms^{L_2',n}(bk) \wedge g_1' {\compare} g_2' \}$ be the set of pair of comparable groups in the matching sets.
	For $(g_1',g_2') {\in} CG(ms^{L_1',n}(bk),ms^{L_2',n}(bk))$, $g_2'$ has crack size $cs=|g_2'|-min(|g_1'|,|g_2'|)$. 
	$C(ms^{L_1',n}(bk),ms^{L_2',n}(bk))=\sum cs$ is the number of excludable cases from $ms^{L_1',n}(bk)$ exploiting the $C$-attack, where $\sum$ is over $(g_1',g_2') {\in} CG(ms^{L_1',n}(bk),ms^{L_2',n}(bk))$.
	
\end{theorem}
\vspace{-0.5 cm}
\begin{proof}
	Consider $(g_1',g_2') {\in} CG(ms^{L_1',n}(bk),ms^{L_2',n}(bk))$. Based on \autoref{lemma:buddy_max}, if ${|g_2'| > |g_1'|}$, at least $|g_2'|-min(|g_1'|,|g_2'|)$ process instances in $g_2'$ do not have a buddy in $g_1'$ for any linker. Such process instances either cannot match the given BK, according to \autoref{lemma:compare_2}, or they have been started at timestamp $t$, $t_1{<}t{\leq}t_2$.
\end{proof}

\begin{exmp}
	Consider $L_1'$ and $L_2'$ in Figure~$\ref{fig:anon}$, $n{=}1$, and $bk{=}\langle d,e \rangle$. $|g_1'|{=}2$ and $|g_2'|{=}3$ for the HIV groups in $ms^{L_1',n}(bk)$ and $ms^{L_2',n}(bk)$, respectively. $cs{=}3-min(2,3)$ is the crack size of $ms^{L_2',n}(bk)$ based on $C$-attack.
\end{exmp}

\begin{definition}[$C$-Anonymity]
	\label{def:c-anonymity}
	Let $L_1'$ and $L_2'$ be two anonymized event logs at $t_1$ and $t_2$, and $n {\in} \mathbb{N}_{\ge 1}$ be the anonymization parameter. 
	The $C$-anonymity of $L_1'$ and $L_2'$ is ${CA}^{n}(L_1'{,}L_2')=\min\limits_{bk \in \pactivities^*}|ms^{L_2',n}(bk)|-C(ms^{L_1',n}(bk),ms^{L_2',n}(bk))$.
\end{definition}

\begin{lemma}
	\label{lemma:b-anonymity}
	Let $L_1'$ and $L_2'$ be two anonymized event logs released at timestamps $t_1$ and $t_2$, $bk \in \pactivities^*$ be the BK, and $n {\in} \mathbb{N}_{\ge 1}$ be the anonymization parameter.
	Consider $g_2' {\subseteq} ms^{L_2',n}(bk)$, $G_1' {=} \{ p_1' {\in} L_1' \mid \exists_{p_2' \in g_2'} p_1' {\compare} p_2'\}$, and $G_2' {=} \{ p_2' {\in} L_2' \mid \exists_{p_1' \in G_1'} p_1' {\compare} p_2'\}$. Every process instance in $G_2'$ is comparable to all records in $G_1'$ and only those records in $G_1'$.
\end{lemma}

\begin{theorem}[Crack size based on $B$-attack]
	\label{theory:B-anonimity}
	Let $bk {\in} \pactivities^*$ be the BK, $n {\in} \mathbb{N}_{\ge 1}$ be the anonymization parameter, and $L_1'$ and $L_2'$ be two anonymized event logs released at timestamps $t_1$ and $t_2$. 
	Let $g_2' {\subseteq} ms^{L_2',n}(bk)$, $G_1' {=} \{ p_1' {\in} L_1' \mid \exists_{p_2' \in g_2'} p_1' {\compare} p_2'\}$, and $G_2' {=} \{ p_2' {\in} L_2' \mid \exists_{p_1' \in G_1'} p_1' {\compare} p_2'\}$.
	$g_2'$ has crack size $cs=max(0,|G_1'|-(|G_2'|-|g_2'|))$.  
	$B(ms^{L_2',n}(bk),L_1',L_2')=\sum_{g_2' \in ms^{L_2',n}(bk)} cs$ is the number of excludable cases from $ms^{L_2',n}(bk)$ exploiting $B$-attack.
\end{theorem}
\vspace{-0.4 cm}
\begin{proof}
	\label{proof:b-anonymity}
	According to \autoref{lemma:b-anonymity}, all process instances in $G_1'$ and only those process instances can have a buddy in $G_2'$. Therefore, each process instance in $G_1'$ has a buddy either in $g_2'$ or $G_2' - g_2'$. If $|G_1'|>|G_2'|-|g_2'|$, then $|G_1'|-(|G_2'|-|g_2'|)$ process instances in $g_2'$ must be started at timestamp $t$, $t {\le} t_1$.  
\end{proof}

\begin{exmp}
	Consider $L_1'$ and $L_2'$ in Figure~$\ref{fig:anon}$, $n{=}1$, and $bk{=}\langle d,c \rangle$. $|g_2'|{=}3$ for the Corona group in $ms^{L_2',n}(bk)$, $G_1'{=}\{ p_1' {\in} L_1' \mid \pi_c(p_1') {\in} \{10,20,30\} \}$, and $G_2'=\{ p_2' {\in} L_2' \mid \pi_c(p_2') {\in} \{41,51,81,91,95\} \}$. $cs{=}max(0,3-(5-3))$ is the crack size of $ms^{L_2',n}(bk)$ based on $B$-attack.
\end{exmp}

\begin{definition}[$B$-Anonymity]
	\label{def:b-anonymity}
	Let $L_1'$ and $L_2'$ be two anonymized event logs at $t_1$ and $t_2$, and $n {\in} \mathbb{N}_{\ge 1}$ be the anonymization parameter. 
	$BA^n(L_1',L_2')=\min\limits_{bk \in \pactivities^*}|ms^{L_2',n}(bk)|-B(ms^{L_2',n}(bk),L_1',L_2')$ is the $B$-anonymity of $L_1'$ and $L_2'$. 
\end{definition}

%

Given $n {\in} \mathbb{N}_{\ge 1}$ as the anonymization parameter, \footnotesize$KA^n(L'){=}\min\limits_{bk \in \pactivities^*} |ms^{L',n}(bk)|$ \normalsize is the $k$-anonymity of an anonymized event log $L'$ w.r.t. $n$. 
Assuming $L_1'$ and $L_2'$ as two anonymized event logs at timestamps $t_1$ and $t_2$, we calculate \textit{the proportion of the cracked cases} (PoCs) after launching the correspondence attacks as follows: \footnotesize $FC^n(L_1',L_2'){=}\frac{(KA^n(L_1'){-}FA^n(L_1',L_2'))}{KA^n(L_1')}$, $CC^n(L_1',L_2'){=}\frac{(KA^n(L_2'){-}CA^n(L_1',L_2'))}{KA^n(L_2')}$, \normalsize and \footnotesize$BC^n(L_1',L_2'){=}\frac{(KA^n(L_2'){-}BA^n(L_1',L_2'))}{KA^n(L_2')}$.

\section{Experiments}\label{sec:experiments}
In this section, we employ \textit{Sepsis} \cite{Sepcis_2016_Felix_short} as a real-life event log and simulate different continuous event data publishing scenarios. We report privacy losses and anonymity values based on the correspondence attacks. Note that \textit{Sepsis} is one of the most challenging event logs for PPTs \cite{pretsaICPM2019_short,rafieitlkc_short,MannhardtKBWM19_short}. We consider two main scenarios to cover various situations w.r.t. \textit{event data volume} and \textit{velocity of event data publishing}. In both scenarios, we consider two releases to be published.

In \textbf{Scenario \RN{1}}, we consider the entire event log as the second collection of events $L_2(100)$. Keeping the second collection of events as $L_2(100)$, we generate four different variants for the first collection of events named $L_1(99)$, $L_1(95)$, $L_1(90)$, and $L_1(75)$, s.t., $L_1(x)$ contains $x\%$ of cases. Note that we ignore the decimal points for percentages, e.g., 90\% could be 90.01\% or 90.95\%.
In \textbf{Scenario \RN{2}}, we filter 50\% of cases as the first collection of events $L_1(50)$. Keeping the first collection of events as $L_1(50)$, we generate four different variants for the second collection of events named $L_2(51)$, $L_2(55)$, $L_2(60)$, and $L_2(75)$, s.t., $L_2(x)$ contains $x\%$ of cases. 
To filter the event logs, we use \textit{time-frame filtering} where the start time is always the start time of the event log and the end time is changed to pick the desired percentage of cases. 

\begin{figure}[t]
	\centering
	\subfloat[\scriptsize The anonymity values when the gap between two releases is $\le$1\%, i.e., $L_1'$ and $L_2'$ were obtained from $L_1(99)$ and $L_2(100)$, respectively. ]{\includegraphics[width=0.49\textwidth]{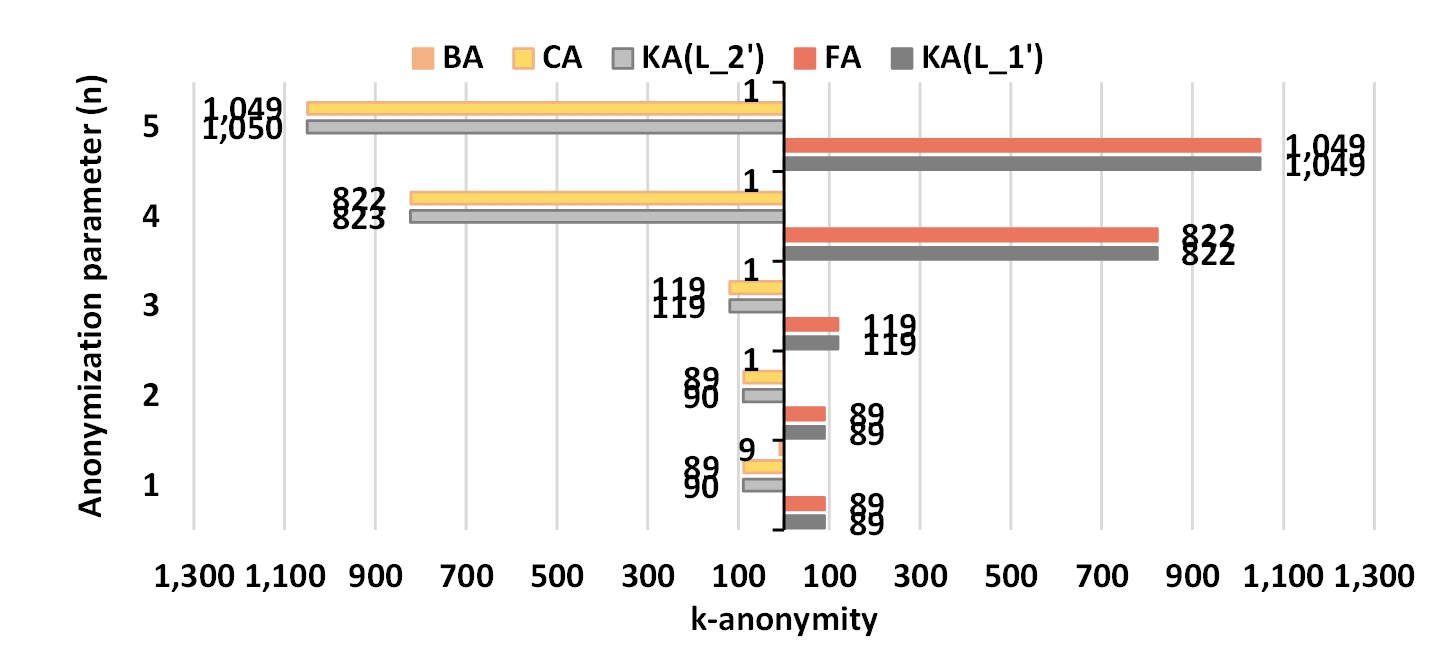}\label{fig:sc1_value_1}}
	\hfill
	\subfloat[\scriptsize The anonymity values when the gap between two releases is $\le$5\%, i.e., $L_1'$ and $L_2'$ were obtained from $L_1(95)$ and $L_2(100)$, respectively.
	]{\includegraphics[width=0.49\textwidth]{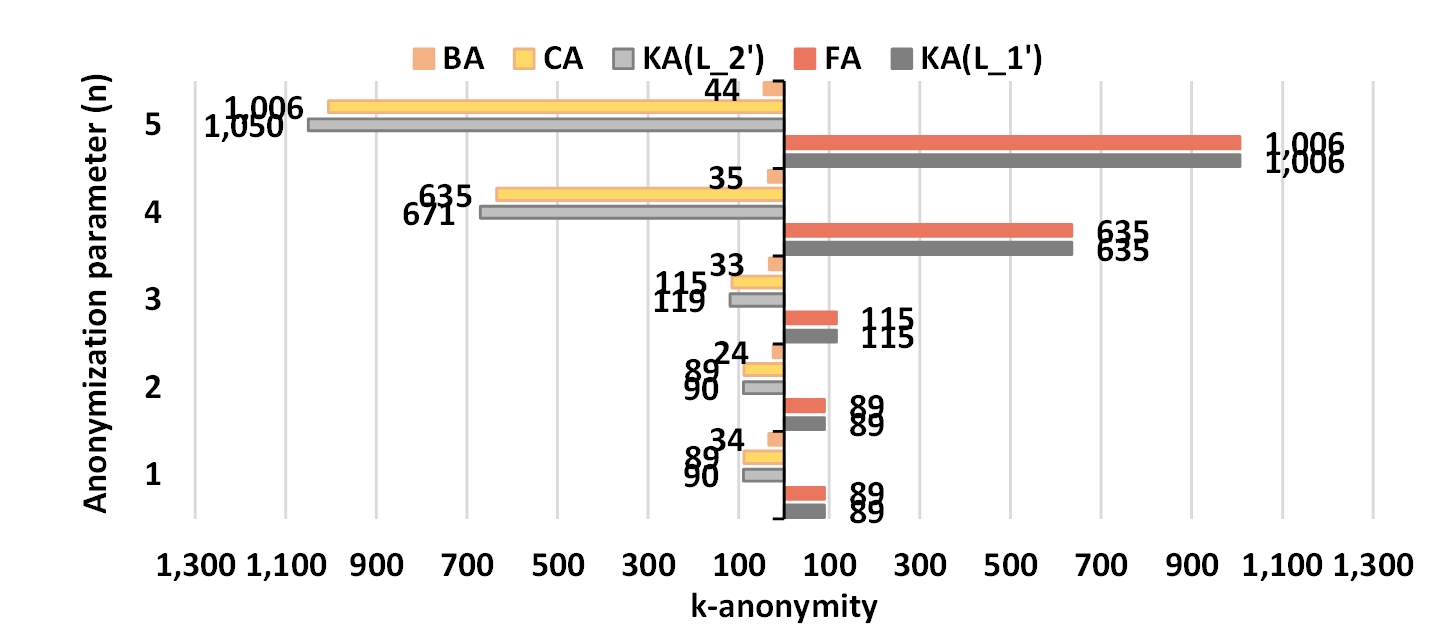}\label{fig:sc1_value_5}} \hfill
	\subfloat[\scriptsize The anonymity values when the gap between two releases is $\le$10\%, i.e., $L_1'$ and $L_2'$ were obtained from $L_1(90)$ and $L_2(100)$, respectively. 
	]{\includegraphics[width=0.49\textwidth]{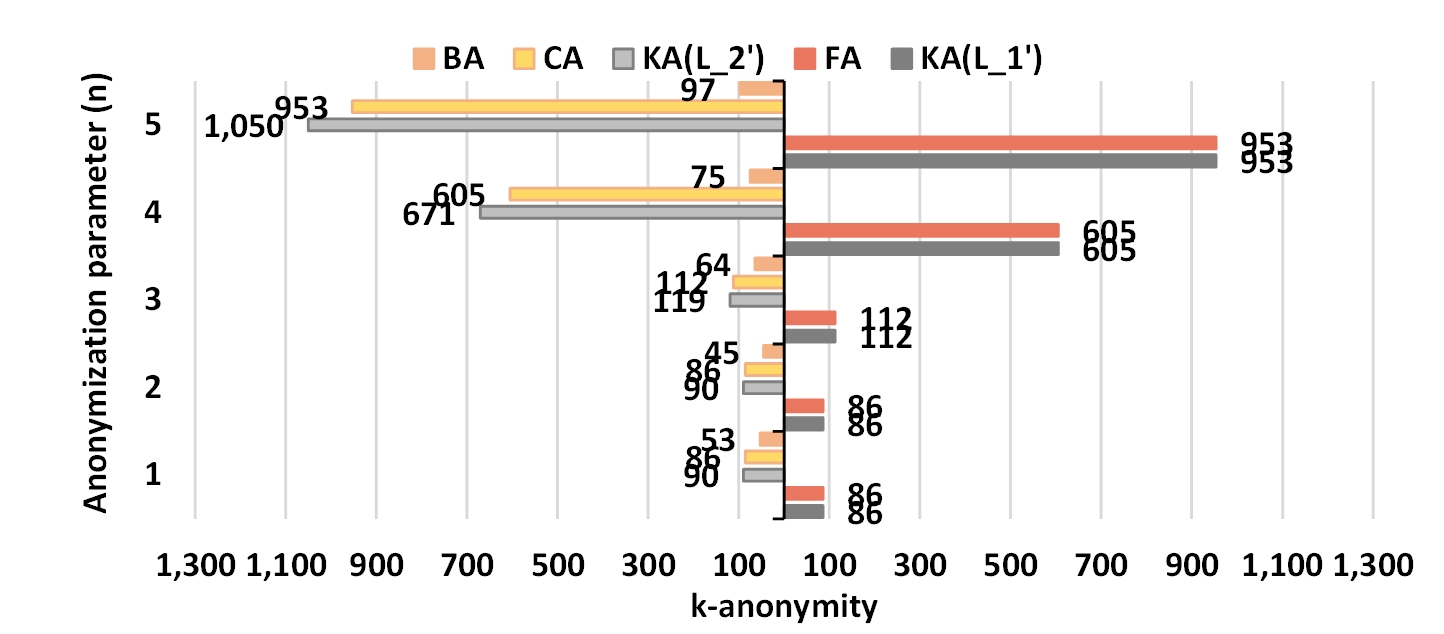}\label{fig:sc1_value_10}} \hfill
	\subfloat[\scriptsize The anonymity values when the gap between two releases is $\le$25\%, i.e., $L_1'$ and $L_2'$ were obtained from $L_1(75)$ and $L_2(100)$, respectively.
	]{\includegraphics[width=0.49\textwidth]{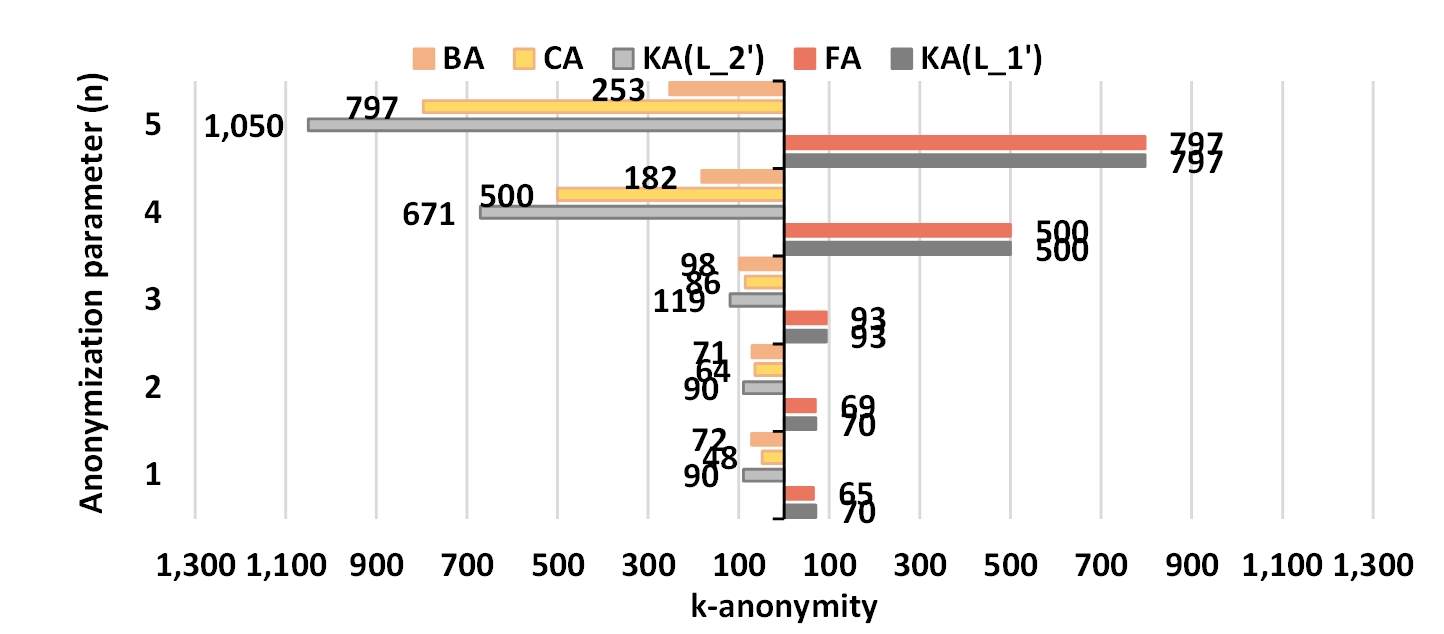}\label{fig:sc1_value_25}} \hfill
	\caption{The anonymity values for different variants of pairs of anonymized releases in Scenario \RN{1}. $KA(L_1')$ is $k$-anonymity of $L_1'$, $KA(L_2')$ is $k$-anonymity of $L_2'$, $FA$ is $k$-anonymity of $L_1'$ after launching $F$-attack, $CA$ is $k$-anonymity of $L_2'$ after launching $C$-attack, and $BA$ is $k$-anonymity of $L_2'$ after launching $B$-attack.}
	\label{fig:scenario_1_values}
\end{figure}

In both scenarios, the gap between two collections varies, s.t., it contains at most 1\%, 5\%, 10\%, or 25\% new cases.  
We focus on the percentage of cases rather than a fixed time window, e.g., daily, weekly, etc., because a fixed time window could contain different amount of data in different slots.
We employ the extended version of TLKC-privacy model \cite{rafieitlkc_short} as the group-based PPT where one can adjust power and type of BK.\footnote{https://github.com/m4jidRafiei/TLKC-Privacy-Ext} 
The model removes events from traces w.r.t. \textit{utility loss} and \textit{privacy gain} to provide the desired privacy requirements. 
We consider all the possible sequences of activities in the event log with the maximal length 5 as the candidates of BK, and $k{=}20$ as the lower bound for $k$-anonymity, i.e., the privacy model guarantees that a single release of the event log meets at least 20-anonymity for all the candidates of BK.  
On the data recipient's side, in each scenario, four different pairs of anonymized releases are received. 
We developed a Python program to detect the attacks and report the anonymity values. The source code and other resources are available on GitHub.\footnote{https://github.com/m4jidRafiei/PP\_CEDP}

\begin{figure}[t]
	\centering
	\includegraphics[width=0.98\textwidth]{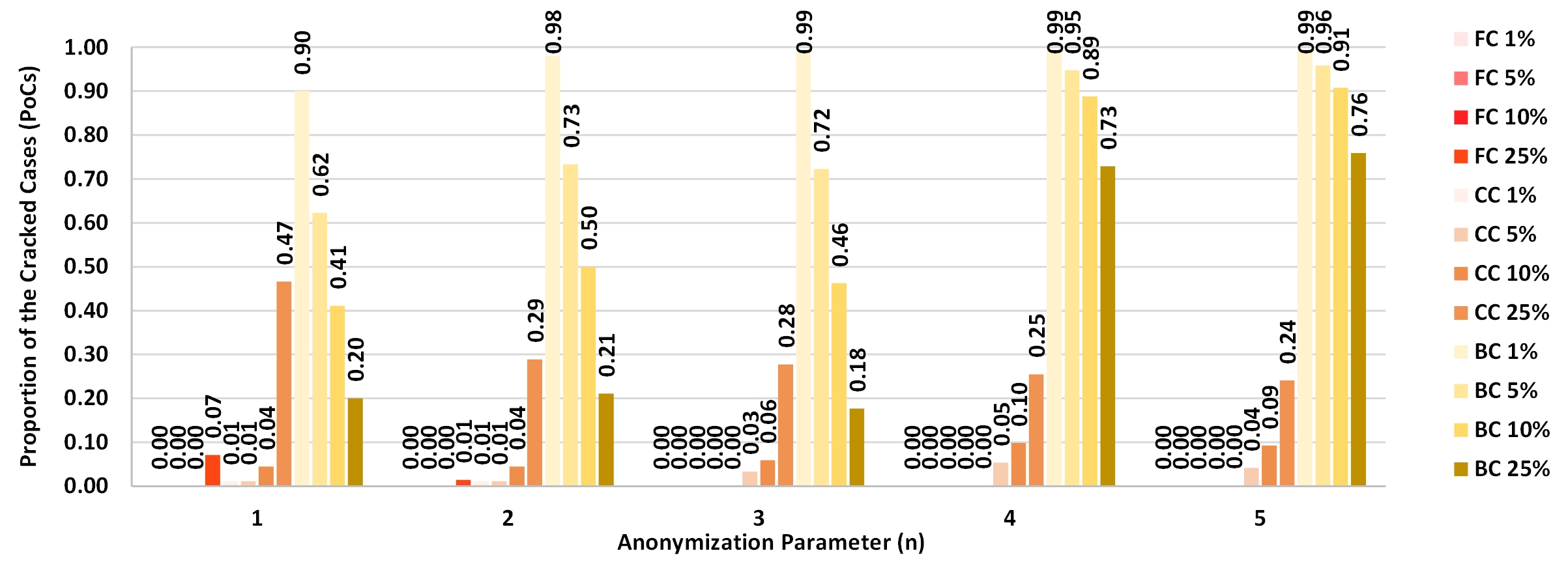}
	\caption{Let x be the maximal gap between two anonymized releases. FC x\%, CC x\%, and BC x\% show the PoCs exploiting $F$-attack, $C$-attack, and $B$-attack, respectively. For each anonymization parameter, the first, the second, and the third 4 bars show the results for $F$-attack, $C$-attack, and $B$-attack, respectively.}\label{fig:sc1_crack_all}
\end{figure}

Figure~\ref{fig:scenario_1_values} shows the anonymity values before and after launching the attacks in Scenario \RN{1}. Note that when $n$ is equal to the length of the BK, all cases already fall into the matching sets. Therefore, the maximal value for the anonymization parameter is 5 which is the maximal length assumed for the BK. 
Figure~\ref{fig:sc1_value_1} shows that when the gap is at most 1\% and $n{=}1$, the anonymized release $L_2'$ has 90-anonymity. However, after launching the $B$-attack, 81 cases are cracked, i.e., 90\% of cases, and $k$-anonymity is degraded to 9, i.e., $BA^1(L_1',L_2'){=}9$. For $n{>}1$, the $B$-anonymity is 1, i.e., there exists a sequence of activities of the maximal length 5 that can be used to uniquely identify a case assuming that at most $n{>}1$ activities have been removed by the PPT. Note that the second release includes only 1 new case when the gap is at most 1\%.     

Figure~\ref{fig:sc1_crack_all} shows how the PoCs are changed when we vary the anonymization parameter $n$ in Scenario \RN{1}. 
Each pair of the anonymized releases is indicated with the percentage of the gap, e.g., 1\% in Scenario \RN{1} indicates two releases obtained from $L_1(99)$ and $L_2(100)$.
When the gap between two releases is small, the $B$-attack results in much higher values for the PoCs compared to the other attacks. However, when the gap becomes larger, the PoCs of the $B$-attack decreases.
This happens because for the smaller $L_1'$s, there exist fewer cases that can be excluded from the matching sets in $L_2'$ because of their timestamps. 
The $C$-attack shows different behavior that is due to the assumed timestamp for the victim case, i.e., for the larger gaps, there exist more cases that their timestamps comply with the second release $L_2'$ and cannot have a corresponding case in $L_1'$.
The $F$-attack cracks fewer cases, which is expected because its target release is $L_1'$, and it only exploits the BK mismatching.
Note that greater values for the anonymization parameter mean that the adversary assumes higher data distortion which results in greater values for the anonymity. 
We had similar observations for Scenario \RN{2}, and the results are available in our GitHub repository.

\section{Extensions}\label{sec:extensions}
The two releases scenario can be extended to the general scenario where more releases are involved. In the general scenario, we consider $m {\in} \mathbb{N}_{> 2}$ collections of events $L_1,L_2,\dots,L_m$ collected at timestamps $t_1,t_2,\dots,t_m$ and published as $L_1',L_2',\dots,L_m'$. 
The correspondence knowledge is also extended, s.t., every case in $L_i'$ has a corresponding case in $L_j'$, $i{<}j{\le} m$. Consider the introduced attacks based on two releases as \textit{micro attacks}. Given more than two releases, the adversary can launch two other types of attacks, so-called \textit{optimal micro attacks} and \textit{composition of micro attacks} \cite{fungCDP}.

\textbf{Optimal micro attacks:} The idea is to find the best background release which results in the largest possible crack size. For instance, consider the $F$-attack on $L_i'$. The adversary can choose any $L_j'$, $i{<}j{\le} m$, as the background release. Let $bk \in \pactivities^*$ be the background knowledge, $n {\in} \mathbb{N}_{\ge 1}$ be the anonymization parameter, and $cs_{ij}$ be the crack size of a pair of comparable groups $(g_i',g_j') \in CG(ms^{L_i',n}(bk),ms^{L_j',n}(bk))$. The optimal crack size of $g_i'$ is $\max\limits_{i{<}j{\le} m} cs_{ij}$.

\textbf{Composition of micro attacks:} The idea is to compose multiple micro attacks to increase the crack size of a group. The micro attacks are launched one after the other. Note that the composition is not possible for any arbitrary choice of micro attacks. It is possible only if all the micro attacks in the composition assume the same timestamp for the victim case, and the required correspondence knowledge holds for the next attack after the previous attack \cite{fungCDP}. Hence, considering $L_i'$, $L_j'$, and $L_l'$, as the anonymized releases, s.t., $i{<}j{<}l{\leq}m$, only two compositions are possible: (1) $B$-attack on $L_i'$ and $L_j'$ followed by $F$-attack on $L_j'$ and $L_l'$, and (2) $B$-attack on $L_i'$ and $L_j'$ followed by $C$-attack on $L_j'$ and $L_l'$. 

Here, we focused on $k$-anonymity which is the foundation for the group-based PPTs. The proposed approach can be extended to cover all the extensions of $k$-anonymity introduced to deal with \textit{attribute linkage} attacks, e.g., $l$-diversity, $(\alpha,k)$-anonymity, confidence bounding, etc. The measures of such PPTs can be modified to consider the cracked cases. Moreover, new group-based PPTs for process mining can be designed to consider $F/C/B$-anonymity. 
For example, a naive algorithm is to start with the maximal possible anonymity, i.e., having only one trace variant, e.g., the longest common subsequence, and then adding events w.r.t. their effect on data utility and privacy loss.        

\section{Related Work}\label{sec:related_work}
Privacy/confidentiality in process mining is growing in importance. The work having been done covers different aspects of the topic including \textit{the challenges} \cite{van2016responsible_short,mannhardt2018privacy_short,pika2020privacy_short}, \textit{confidentiality frameworks} \cite{rafieiWA19_short}, \textit{privacy by design} \cite{MichaelKMBR19_short}, \textit{privacy guarantees} \cite{MannhardtKBWM19_short,pripel_short,rafieitlkc_short,pretsaICPM2019_short}, \textit{inter-organizational privacy issues} \cite{smcProcessMining_short}, and \textit{privacy quantification} \cite{riskProcessMining_short,rafiei_quantification}.   	
Confidentiality is one of the important challenges of the bigger sub-discipline of process mining called \textit{Responsible Process Mining} (RPM) \cite{van2016responsible_short}.
In \cite{mannhardt2018privacy_short}, the authors provide an overview of privacy challenges for process mining in human-centered industrial environments.
In \cite{pika2020privacy_short}, the authors focus on data privacy and utility requirements for healthcare event data. 
A general framework for confidentiality in process mining is proposed in \cite{rafieiWA19_short}. 
In \cite{MichaelKMBR19_short}, the goal is to propose a privacy-preserving system design for process mining. 
In \cite{rafiei2019role_short}, the authors introduce a privacy-preserving method for discovering roles from event logs.
In \cite{pretsaICPM2019_short}, $k$-anonymity and $t$-closeness are adopted to preserve the privacy of \textit{resources} in event logs. 
In \cite{MannhardtKBWM19_short,pripel_short}, the notion of \textit{differential privacy} is utilized to provide privacy guarantees.
In \cite{rafieitlkc_short}, the TLKC-privacy is introduced to deal with high variability issues in event logs for applying group-based anonymization techniques.
A secure multi-party computation solution is proposed in \cite{smcProcessMining_short} for preserving privacy in an inter-organizational setting.
In \cite{riskProcessMining_short}, the authors propose a measure to evaluate the re-identification risk of event logs. Also, in \cite{rafiei_quantification}, a general privacy quantification framework, and some measures are introduced to evaluate the effectiveness of PPTs.
In \cite{rafieippdp_arxiv}, the authors propose a privacy extension for the XES standard to manage privacy metadata.


\section{Conclusion}\label{sec:conclusion}
In practice, event data need to be published continuously to keep the process mining results up-to-date. In this paper, for the first time, we focused on the attacks appearing when anonymized event data are published continuously. 
We formalized three different types of the so-called \textit{correspondence attacks} in the context of process mining: $F$-attack, $C$-attack, and $B$-attack. We demonstrated the attack detection techniques to quantify the anonymity of event logs published continuously. 
We simulated the continuous event data publishing for real-life event logs using various scenarios. For an example event log, we showed that the provided privacy guarantees can be degraded exploiting the attacks.
The attack analysis and detection techniques can be adjusted and attached to different group-based PPTs to enhance the privacy guarantees when event data are published continuously. 
In this paper, we mainly focused on \textit{suppression} as the anonymization operation. In future, other anonymization operations such as \textit{addition} or \textit{swapping} could be analyzed.
Similar attack analysis can be done for other types of PPTs, e.g., \textit{differential privacy}, in the context of process mining to protect provided privacy guarantees. Moreover, one could evaluate the effect of continuous publishing scenarios on privatized process mining results.   

\section*{Acknowledgment} Funded under the Excellence Strategy of the Federal Government and the L{\"a}nder. We also thank the Alexander von Humboldt Stiftung for supporting our research.

\bibliographystyle{splncs04}
\bibliography{Refrences}

\end{document}